# Dynamics of Voltage Driven Self-Sustained Oscillations in NdNiO$_3$ Neuristors

Upanya Khandelwal[1], Qikai Guo[2], Beatriz Noheda[2,3], Pavan Nukala[1*], Saurabh Chandorkar[1*]

[1] *Center for Nanoscience and Engineering, Indian Institute of Science, Bengaluru, Karnataka, India, 560012*
[2] *Faculty of Science and Engineering, University of Groningen, Groningen, Netherlands 9747 AG*
[3] *Cognigron University of Groningen, Groningen, Netherlands 9747 AG*

**Corresponding authors:**

Saurabh Chandorkar: saurabhc@iisc.ac.in

Pavan Nukala : pnukala@iisc.ac.in

**ABSTRACT:**

Active memristor elements, also called neuristors, are self-oscillating devices that are very good approximations to biological neuronal functionality and are crucial to the development of low-power neuromorphic hardware. Materials that show conduction mechanisms that depend superlinearly with temperature can lead to negative differential resistance (NDR) regimes, which may further be engineered as self-oscillators. Thermal runaway, insulator to metal phase transitions (IMT) can lead to such superlinearity and are being extensively studied in systems such as TaOx, NbOx and VO$_2$. However, ReNiO$_3$ systems that offer large tunability in metal-insulator transition temperatures are less explored so far. Here we demonstrate all-or-nothing neuron-like self-oscillations at MHz frequency and low temperatures on thin films of NdNiO$_3$, a model charge transfer insulator, and their frequency coding behavior. We study the temperature dependence of NDR and show that it vanishes even at temperatures below the IMT temperature. We also show that the threshold voltages scale with device size and that a simple electrothermal device model captures all these salient features. In contrast to existing models, our model correctly predicts the independence of oscillation amplitude with the applied voltage, offering crucial insights about the nature of fixed points in the NDR region, and the dynamics of non-linear oscillations about them.

1. INTRODUCTION

Current-controlled negative differential resistance (CC-NDR) in strongly correlated [1]oxide based devices, such as niobium dioxide[2–5] and vanadium dioxide[6] has recently gained a lot of attraction in the context of neuromorphic computing[7,8]. CC-NDR is associated with conduction mechanisms that show superlinear dependence on temperature[9], typically either driven by thermal runaway effects or insulator to metal transition (IMT) triggered by Joule heating. [10,11]. Quasistatic NDR results from device instability and local activity[12,13], and under appropriate conditions can lead to electrical self-oscillations[14–17]. Devices displaying an NDR regime are able to amplify electrical signals, within a given parameter range, and are referred to as locally active[18]. Utilizing such locally active memristors as neuronal elements in a neuromorphic architecture enables low power computing[19–22].

NDR is most often reported in $VO_2$ and $NbO_2$ devices. The IMT transition temperature of $VO_2$ is 340K[23], while that of $NbO_2$ occurs at very high temperatures (~1080 K)[24]. It has recently been put forward that reaching the IMT in a device setting in $NbO_2$ is unlikely and the observed oscillations are most likely due to thermal run-away effects[25–27] that occur at lower temperatures[7,28]. Pervoskite rare- earth nickelates ($ReNiO_3$, Re = Pr, Nd, Sm, Eu,.. Lu), are charge-transfer insulators showing rich correlated electron physics, and exhibit IMT owing to charge disproportionation[29]. Suitable choice of the Re element (or combinations of Re elements) allows for tunability in the transition temperatures between 100 and 800 K [29]. As a result, these systems are suitable playgrounds to independently assess the CC-NDR driven oscillatory behavior originating from IMT, thermal runaway and their coupled effects.

NDR arising from IMT has been recently shown in thin films of $NdNiO_3$ and $SmNiO_3$ [30]. In another work, H-doped $NdNiO_3$ (NNO) was proposed as a potential candidate for memristor based neural network with applications in artificial intelligence, although the relevant effects arise owing to the behavior of hydrogen dopants and not the IMT behavior[31]. However, studies that explore electrical self-oscillations and their dynamics from $ReNiO_3$ based systems are lacking or very exploratory.

In this work, we systematically study the volatile switching behavior, and for the first time determine characteristics of self-sustained electrical oscillations on a model system consisting of NNO films epitaxially grown on $LaAlO_3$ (LAO) substrates. We fabricate lateral two-terminal devices of various dimensions and investigate the temperature and channel length dependency of the CC-NDR behavior. We demonstrate non-linear current oscillations at ~MHz frequencies (and higher harmonics) taking place about a fixed point in the NDR region. The operating point of the dynamical system is set by a suitable choice of biasing resistor and external voltage. We show that the oscillation frequency can be tuned with external voltage, and we gain an accurate and clear understanding of all these salient features through a simple coupled electro-thermal modeling.

## 2. RESULTS AND DISCUSSIONS

### 2.1. Film growth and quality checks:

Epitaxial thin films of NNO (5 nm) were grown on LAO substrates using pulsed laser deposition, with conditions reported elsewhere[32]. The defect density in these films has been systematically controlled[33] and the films used in the present work are very high-quality with low defect densities, as reflected from the structural and electrical property measurements reported in ref [32]. The resistance-temperature measurements reveal a hysteretic first order IMT phase transition, with IMT Temperature ($T_{IMT}$=120 K) (during heating cycle).

### 2.2. Transport characteristics of devices:

On these films we fabricate two terminal lateral devices, with channel lengths[1] 200-800 nm, using e-beam lithography followed by metallization with Pt electrodes (see Fig.1a, Methods). Voltage-controlled (VC) current-voltage (I-V) sweeps (Fig.1b) on these two-terminal devices shows threshold-switching behavior similar to what is reported in[30]. The threshold voltage, defined as the voltage at which the insulating state transforms to a metallic state, decreases with increasing ambient temperature (Fig.1c), as expected (less power is required to Joule-heat the device to $T_{IMT}$ and beyond). We also note that the hysteresis associated with this phase transition reduces with increasing ambient temperature. (Fig.1c). Current-controlled sweeps allow us to access the NDR regions (Fig.1b), during the transition from the insulator to the metallic phase. The device voltage at the onset of NDR($V_N$) (corresponds to threshold voltage in VC-IV measurements) scales with the channel length (Fig.1d).

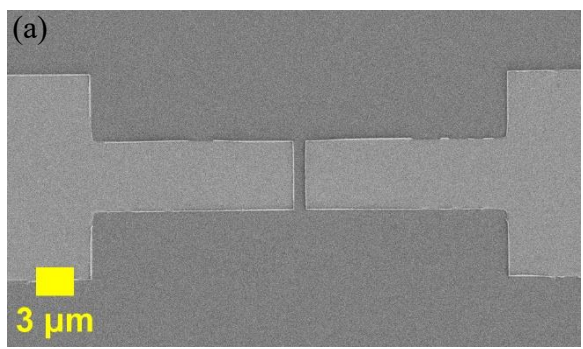
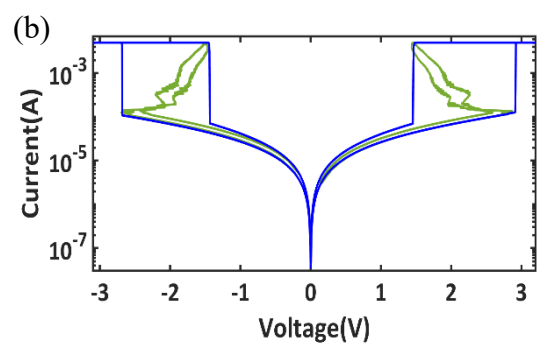

---

[1] The active region between the electrodes will be referred to as a channel.

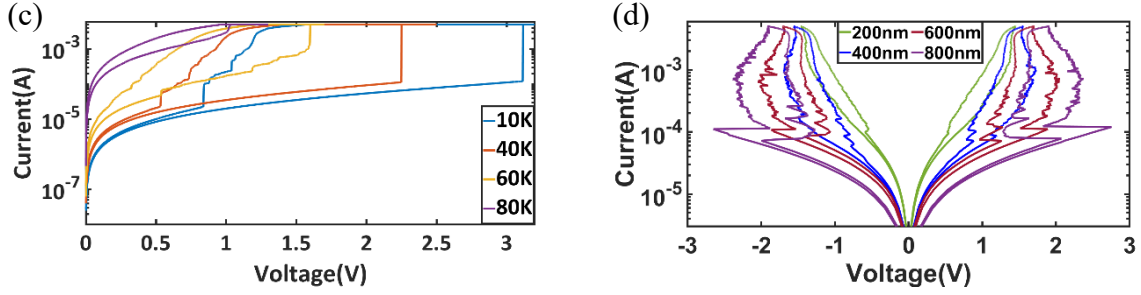

**Fig.1**: DC-I-V characteristics variations with device geometry and temperature (a) Scanning electron microscopy image of one NNO device. (b) I-V characteristics of NNO; blue: voltage-control, green: current-control. (c) Voltage-control I-V at different temperatures. (d) Current-control I-V of various devices with different channel length.

### 2.3. Characteristics and salient features of self-oscillations:

To study the self-oscillations, we added a biasing resistor (10k ohm) in series to the device, and systematically increased the magnitude of voltage pulses (20 µs) from 5V to 15 V while simultaneously measuring the current response. The data on a representative device (channel length=760nm) is shown in Fig.2. For applied voltages below 9V, we observe a regular charging and discharging behavior of the RC circuit element (Fig.2a). However, from 9V to 13V, the device shows periodic non-linear (multiple harmonics), asymmetric current oscillations (Fig.2b,2d). The amplitude of current oscillations remains constant at ~3.5 mA, independent of applied voltage, and this aligns with the all-or-none law in neurons[34,35].

From the load line analysis, we show that oscillations occur only when the device operating point falls in the NDR region (Fig.3a, applied voltages: 9V to 12V). Importantly, we note that the frequency of oscillations increases with the applied voltage, and that its rate of increase, decreases with the resistance of the external resistor. In other words, the frequency of oscillations in our devices encodes information about the applied external stimulus, a behavior known as frequency coding in the nervous system [36]. Furthermore, we note that at a particular voltage, the frequency of oscillations decreases with an increase in resistance of the external resistor (Fig.3b), as also demonstrated in other neuristor systems[37].

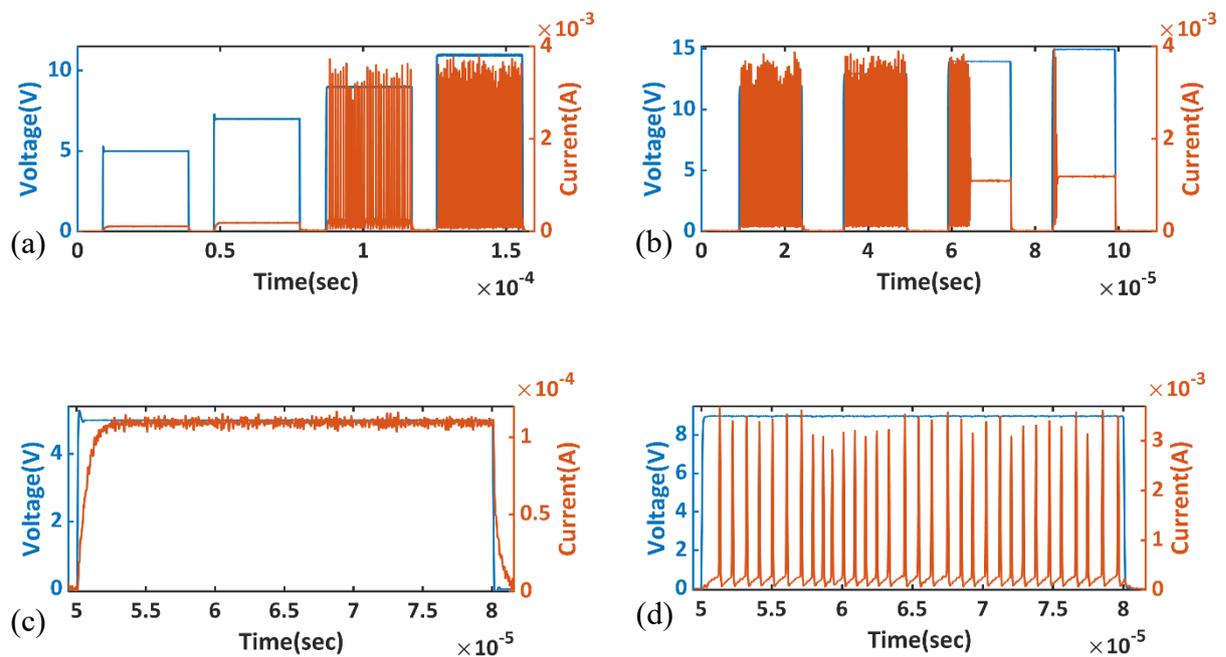

**Fig.2**: Current oscillations. (a) y-axis left: applied voltage (5 to 11V) y-axis right: measured current. (b) y-axis left: applied voltage (12 to 15V) y-axis right: measured current. (c) zoomed-in image of (a) at 5V.(d) zoomed-in image of (a) at 9V.

### 2.4. Electrothermal device modeling:

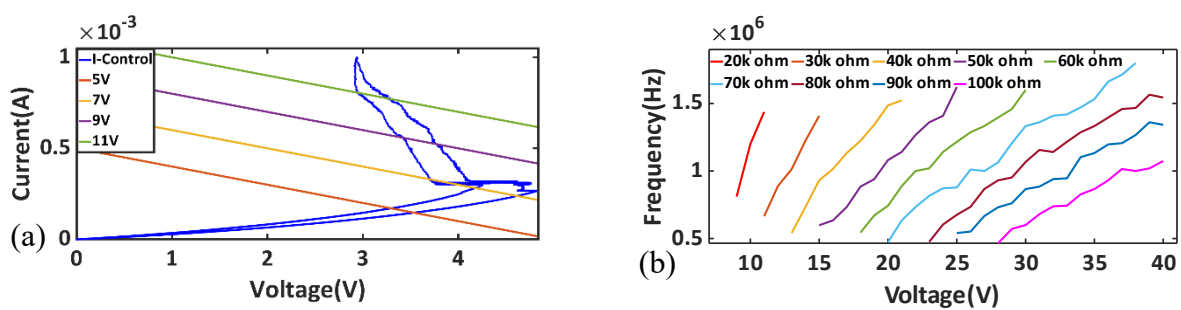

**Fig.3**: Load-line analysis and frequency coding (a) Load-line analysis for 10k ohm resistor. (b) Frequency of current oscillations with voltage for different biasing resistors.

To understand the various features of the observed CC-NDR and self-sustained oscillatory behavior, we carried out electrothermal device modeling using LT Spice. The coupled electrical and heat transport problem is represented by the circuits shown in Figs. S1-S3.

The electrical model for our device consists of resistors $R_x$, $R_d$ and a device capacitance $C_d$ in a parallel configuration. An appropriate source can be applied across the device to match the measurement protocols and a source and line capacitance has also been included to accurately model the physics of the system. It has been shown in other works[38,39] that only a part of the channel participates in the phase transition, which is modeled here as a non-linear resistor ($R_d$) that changes with device temperature as per the conductivity vs temperature characteristics[32]. $R_x$, on the other hand is the background resistance of the non-active region in the channel and is assumed to be relatively independent of temperature as substantiated by our thermal model. Incorporation of a separate $R_x$ allows us to correctly predict the order of magnitude of the values of currents in self-oscillations. The output power of the electrical circuit is then used as the heat source for the thermal circuit. Heat can be dissipated vertically across various layers to the substrate, or laterally along the length of the device and along the width outside the channel. Various heat dissipation channels are lumped as thermal resistors, with equivalent thermal resistance ($R_t$) given in Eq.1. Equivalent thermal capacitances ($C_t$) of various layers are given in Eq.1.

The thermal resistance ($R_t$) and thermal capacitance ($C_t$) are estimated from the material parameters and geometry as follows:

$$R_t = \frac{L}{kA} , C_t = \rho C V \qquad (Eq.1)$$

where, L: length of the heating element, k: thermal conductivity of material, A: cross-sectional area, $\rho$: mass density, C: heat capacity and V: volume. (Details in supplementary).

Thermal and electrical conductivities of the device as a function of temperature across the phase transition in NNO are modeled as a sigmoidal function, as shown in equations 2 and 3 (also see Supplementary table for exact values used in the model).

$$\sigma = \sigma_{ins} + (\sigma_{ins} - \sigma_m) \frac{1}{1+\exp\left(\frac{T_s-T}{\alpha}\right)} \qquad (Eq.\ 2)$$

$$\kappa = \kappa_{ins} + (\kappa_{ins} - \kappa_m) \frac{1}{1+\exp\left(\frac{T_s-T}{\alpha}\right)} \qquad (Eq.\ 3)$$

Here, $\sigma_{ins}, \sigma_m$ and $\kappa_{ins}, \kappa_m$ are the insulating and metallic electrical and thermal conductivity respectively. $T_s$ and $\alpha$ are the transition temperature and the spread about the transition temperature respectively. The various results of our electrothermal simulations are shown in Fig.5.

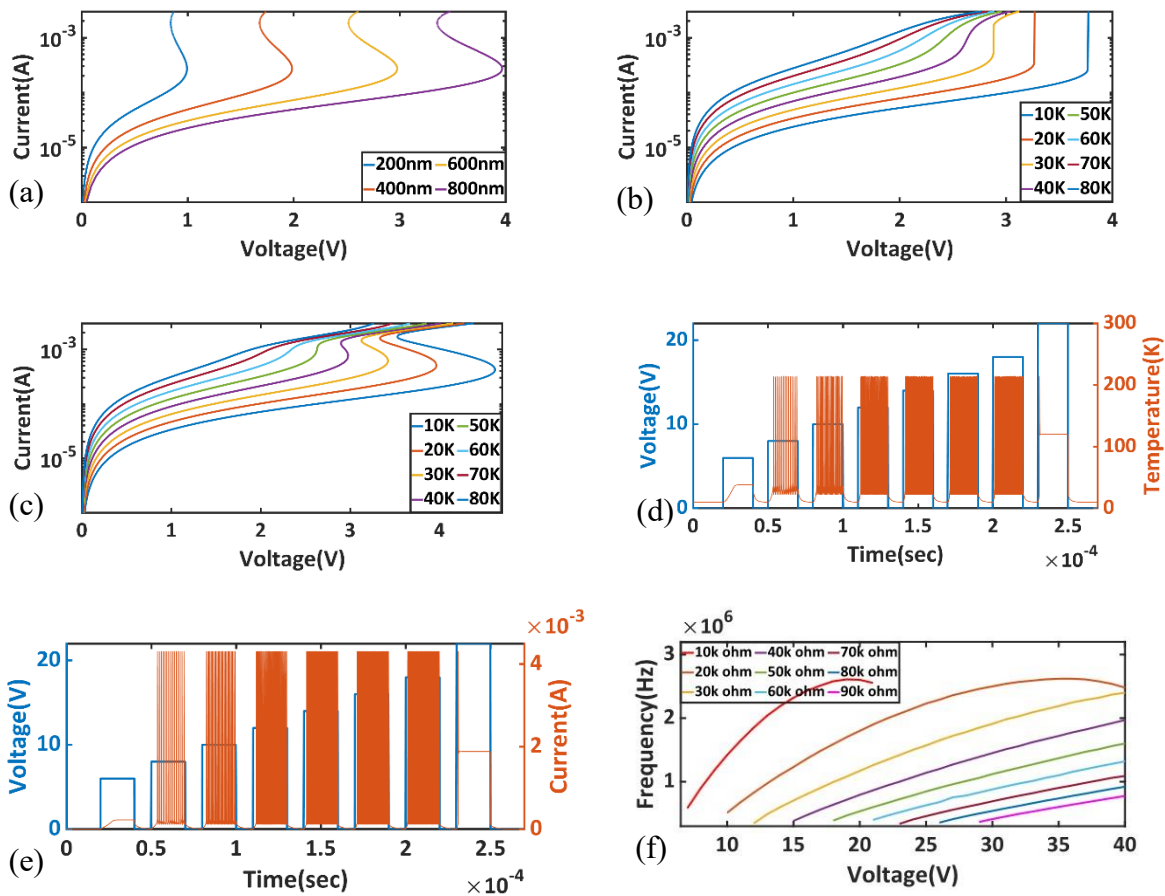

**Fig.4**: Electrothermal simulations. (a) Current-control I-V for different channel lengths at 10K. (b) Voltage-control I-V at different ambient temperatures. (c) Current-control I-V at different ambient temperatures for a 760nm channel length device. (d) Time sequence of: y-axis left, the applied voltage pulses (7V to 22V) and y-axis right, simulated current. (e) y-axis left: applied voltage (7V to 22V) y-axis right: simulated temperature. (f) Frequency of current and temperature oscillations as a function of voltage for different biasing resistors.

Our results accurately reflect the CC-NDR behavior and the reduction in $V_N$ with reduction in channel length (Fig 5a). This is a result of larger heat dissipated in longer channels, which requires them to be compensated with larger power input (or larger $V_N$) to heat the device to $T_{IMT}$. The decrease of threshold voltage with increasing ambient temperature is also nicely captured (Fig. 5b), confirming the predominant role of Joule heating in IMT. Furthermore, our modeling predicts the weakening of NDR with increased ambient temperature (Fig.5c), and the complete disappearance of the same at an ambient temperature of ~60K, which is much below $T_{IMT}$ (120 K). Indeed, we experimentally verify that devices operating at 80K do not exhibit a threshold voltage of transition (Fig.1c) or associated self-sustained oscillations. We also model the dependence of frequency on biasing resistor and applied voltage (Fig.5f). These curves show a non-linear dependence of frequency with voltage. Although our experiments did not reflect this non-linearity clearly, it was observed experimentally in other threshold memristor systems[40].

Next, using the same parameters for the model as that used for simulating I-V characteristics, we simulated the current oscillations for a specific device (l=760nm) at an ambient temperature of 10K, and a biasing resistor (10 kΩ) in series, with a $V_0$ voltage applied across the circuit (Fig.S2). For $V_0$ in the range 8V to 18V, the device local temperature oscillates from 22K to 214K (Fig.5d), independent of the value of $V_0$. This corresponds to a few mA fluctuations in current (Fig.5e), also independent of $V_0$, as observed experimentally. A representative limit cycle for the oscillation of the device is shown in Fig. S4.

The non-dependence of current oscillation amplitude on applied voltage was also observed experimentally by us as well as other researchers on different Mott-insulating systems[39]. However, unlike previous models, our simple electrothermal model precisely captures this behavior. Finally, our model also captures the frequency coding of external stimulus behavior observed experimentally (Fig. 3b). This is consistent with larger input voltage requiring less time for a capacitor to charge to the threshold voltage, thereby increasing the oscillation frequency.[41]

## 3. CONCLUSION:

We systematically explored the volatile switching behavior and characteristics of self-sustained electrical oscillations in the lateral, two-terminal $NdNiO_3$ (NNO) devices that occur about a fixed point in the NDR region of the device. Our devices emulate an all-or-nothing neuronal oscillatory behavior[34,35], and also encode the external stimulus in the frequency of the oscillations (frequency coding). Our electrothermal model accurately reproduces all the observed behavior and shows that a superlinear dependence of conductivity on device temperature at the phase transition causes NDR and coupled current oscillations. More importantly, for the first time, our model correctly shows the independence of current fluctuation amplitude with external stimulus. Although this feature was observed in several other neuristors, it could not be captured through models[38,39]. Furthermore, our model seamlessly captures the I-V characteristics of devices with different lengths as well as their dependence on the ambient temperature. Our work opens up a framework to explore the family of rare earth nickelates which offer tunable transition temperatures and hence a platform to study the interplay between multiple mechanisms contributing to NDRs such as coupling of thermal runoff based and IMT based neuristor behaviors and their possible power advantages.

**ACKNOWLEDGEMENTS:**

This work was partly carried out at Micro and Nano Characterization Facility(MNCF), and National Nanofabrication Centre(NNfC) located at CeNSE, IISc, Bengaluru, and benefitted from all the help and support from the staff. P.N. acknowledges Start-up grant from IISc, Infosys Young Researcher award, and DST-starting research grant SRG/2021/000285. The authors acknowledge funding support from the Ministry of Education (MoE) from Ministry of Electronics and Information Technology (MeitY) and Department of Science and Technology (DST) through NNetRA. BN, QG acknowledge the financial support of the CogniGron research center and the Ubbo Emmius Funds (University of Groningen). SC acknowledges Indian Space Research Organization, Government of India (GoI) under Grant DS_2B13012(2)/41/2018-Sec.2, by the Ministry of Electronics and Information Technology, GoI under 25(2)/2020_ESDA DT.28.05.2020 and Ministry of Human Resource and Development, GoI Grant SR/MHRD_18_0017. SC also acknowledges DRDO JATP: DFTM/02/3125/M/12/MNSST-03.

**REFERENCES:**

(1) Georges, A. Strongly Correlated Electron Materials: Dynamical Mean-Field Theory and Electronic Structure. *AIP Conf. Proc.* **2004**, *715* (1), 3. https://doi.org/10.1063/1.1800733.

(2) Kumar, S.; Wang, Z.; Davila, N.; Kumari, N.; Norris, K. J.; Huang, X.; Strachan, J. P.; Vine, D.; Kilcoyne, A. L. D.; Nishi, Y.; Williams, R. S. Physical Origins of Current and Temperature Controlled Negative Differential Resistances in NbO2. *Nat. Commun.* **2017**, *8* (1), 4–9. https://doi.org/10.1038/s41467-017-00773-4.

(3) Liu, X.; Li, S.; Nandi, S. K.; Venkatachalam, D. K.; Elliman, R. G. Threshold Switching and Electrical Self-Oscillation in Niobium Oxide Films. *J. Appl. Phys.* **2016**, *120* (124102). https://doi.org/10.1063/1.4963288.

(4)     Messaris, I.; Tetzlaff, R.; Ascoli, A.; Williams, R. S.; Kumar, S.; Chua, L. A Simplified Model for a NbO2 Mott Memristor Physical Realization. *Proc. - IEEE Int. Symp. Circuits Syst.* **2020**, *2020-Octob*. https://doi.org/10.1109/iscas45731.2020.9181036.

(5)     Gibson, G. A.; Musunuru, S.; Zhang, J.; Vandenberghe, K.; Lee, J.; Hsieh, C. C.; Jackson, W.; Jeon, Y.; Henze, D.; Li, Z.; Stanley Williams, R. An Accurate Locally Active Memristor Model for S-Type Negative Differential Resistance in NbOx. *Appl. Phys. Lett.* **2016**, *108*(2), 023505. https://doi.org/10.1063/1.4939913.

(6)     Nirantar, S.; Mayes, E.; Rahman, M. A.; Ahmed, T.; Taha, M.; Bhaskaran, M.; Walia, S.; Sriram, S. In Situ Nanostructural Analysis of Volatile Threshold Switching and Non-Volatile Bipolar Resistive Switching in Mixed-Phased a-VOx Asymmetric Crossbars. *Adv. Electron. Mater.* **2019**, *5* (12), 1–9. https://doi.org/10.1002/aelm.201900605.

(7)     Kumar, S.; Strachan, J. P.; Williams, R. S. Chaotic Dynamics in Nanoscale NbO 2 Mott Memristors for Analogue Computing. *Nature* **2017**, *548* (7667), 318–321. https://doi.org/10.1038/nature23307.

(8)     Zhou, G.; Wang, Z.; Sun, B.; Zhou, F.; Sun, L.; Zhao, H.; Hu, X.; Peng, X.; Yan, J.; Wang, H.; Wang, W.; Li, J.; Yan, B.; Kuang, D.; Wang, Y.; Wang, L.; Duan, S. Volatile and Nonvolatile Memristive Devices for Neuromorphic Computing. *Adv. Electron. Mater.* **2022**, *8* (7), 1–33. https://doi.org/10.1002/aelm.202101127.

(9)     Gibson, G. A. Designing Negative Differential Resistance Devices Based on Self-Heating. *Adv. Funct. Mater.* **2018**, *28* (22), 1–9. https://doi.org/10.1002/adfm.201704175.

(10)    Kumar, S.; Pickett, M. D.; Strachan, J. P.; Gibson, G.; Nishi, Y.; Williams, R. S. Local Temperature Redistribution and Structural Transition during Joule-Heating-Driven Conductance Switching in VO2. *Adv. Mater.* **2013**, *25* (42), 6128–6132. https://doi.org/10.1002/adma.201302046.

(11)    Olin, S. W.; Razek, S. A.; Piper, L. F. J.; Lee, W. C. Mechanism of the Resistivity Switching Induced by the Joule Heating in Crystalline NbO2. *Adv. Quantum Technol.* **2022**, No. Mc, 1–9. https://doi.org/10.1002/qute.202200067.

(12)    Wang, G.; Wang, F.; Forti, M.; Ascoli, A.; Demirkol, A. S.; Tetzlaff, R.; Slesazeck, S.; Mikolajick, T.; Chua, L. O. On Local Activity and Edge of Chaos in a NaMLab Memristor. *Front. Neurosci. | www.frontiersin.org* **2021**, *1*, 651452.

https://doi.org/10.3389/fnins.2021.651452.

(13) Chua, L. O. Local Activity Is the Origin of Complexity. *Int. J. Bifurcat. Chaos* **2005**, *15* (11), 3435–3456. https://doi.org/10.1142/S0218127405014337.

(14) Kim, H. T.; Kim, B. J.; Choi, S.; Chae, B. G.; Lee, Y. W.; Driscoll, T.; Qazilbash, M. M.; Basov, D. N. Electrical Oscillations Induced by the Metal-Insulator Transition in VO2. *J. Appl. Phys.* **2010**, *107* (023702). https://doi.org/10.1063/1.3275575.

(15) Yi, W.; Tsang, K. K.; Lam, S. K.; Bai, X.; Crowell, J. A.; Flores, E. A. Biological Plausibility and Stochasticity in Scalable VO 2 Active Memristor Neurons. *Nat. Commun.* **2018**, *9* (1). https://doi.org/10.1038/s41467-018-07052-w.

(16) Driscoll, T.; Quinn, J.; Di Ventra, M.; Basov, D. N.; Seo, G.; Lee, Y. W.; Kim, H. T.; Smith, D. R. Current Oscillations in Vanadium Dioxide: Evidence for Electrically Triggered Percolation Avalanches. *Phys. Rev. B - Condens. Matter Mater. Phys.* **2012**, *86* (9), 1–8. https://doi.org/10.1103/PhysRevB.86.094203.

(17) Beaumont, A.; Leroy, J.; Orlianges, J. C.; Crunteanu, A. Current-Induced Electrical Self-Oscillations across out-of-Plane Threshold Switches Based on VO2 Layers Integrated in Crossbars Geometry. *J. Appl. Phys.* **2014**, *115* (154502). https://doi.org/10.1063/1.4871543.

(18) Mainzer, K. Local Activity Principle: The Cause of Complexity and Symmetry Breaking. *Chaos, CNN, Memristors Beyond A Festschrift Leon Chua* **2013**, 146–159. https://doi.org/10.1142/9789814434805_0012.

(19) Yuan, R.; Duan, Q.; Tiw, P. J.; Li, G.; Xiao, Z.; Jing, Z.; Yang, K.; Liu, C.; Ge, C.; Huang, R.; Yang, Y. A Calibratable Sensory Neuron Based on Epitaxial VO2 for Spike-Based Neuromorphic Multisensory System. *Nat. Commun.* **2022**, *13* (1), 1–12. https://doi.org/10.1038/s41467-022-31747-w.

(20) Kumar, S.; Williams, R. S.; Wang, Z. Third-Order Nanocircuit Elements for Neuromorphic Engineering. *Nature* **2020**, *585*, 518–523. https://doi.org/10.1038/s41586-020-2735-5.

(21) del Valle, J.; Ramírez, J. G.; Rozenberg, M. J.; Schuller, I. K. Challenges in Materials and Devices for Resistive-Switching-Based Neuromorphic Computing. *J. Appl. Phys.* **2018**, *124* (21). https://doi.org/10.1063/1.5047800.

(22) del Valle, J.; Salev, P.; Tesler, F.; Vargas, N. M.; Kalcheim, Y.; Wang, P.; Trastoy, J.; Lee, M. H.; Kassabian, G.; Ramírez, J. G.; Rozenberg, M. J.; Schuller, I. K. Subthreshold

Firing in Mott Nanodevices. *Nature* **2019**, *569*, 388–392. https://doi.org/10.1038/s41586-019-1159-6.

(23) del Valle, J.; Vargas, N. M.; Rocco, R.; Salev, P.; Kalcheim, Y.; Lapa, P. N.; Adda, C.; Lee, M. H.; Wang, P. Y.; Fratino, L.; Rozenberg, M. J.; Schuller, I. K. Spatiotemporal Characterization of the Field-Induced Insulator-to-Metal Transition. *Science,* **2021**, *373* (6557), 907–911. https://doi.org/10.1126/SCIENCE.ABD9088/SUPPL_FILE/SCIENCE.ABD9088_SM.PDF.

(24) Shibuya, K.; Sawa, A. Epitaxial Growth and Polarized Raman Scattering of Niobium Dioxide Films. *AIP Adv.* **2022**, *12* (055103). https://doi.org/10.1063/5.0087610.

(25) Karpov, V. G.; Niraula, D. Resistive Switching in Nano-Structures. *Sci. Rep.* **2018**, *8* (1), 1–10. https://doi.org/10.1038/s41598-018-30700-6.

(26) Goodwill, J. M.; Sharma, A. A.; Li, D.; Bain, J. A.; Skowronski, M. Electro-Thermal Model of Threshold Switching in TaOx-Based Devices. *ACS Applied Materials and Interfaces*. **2017**,9,13, 11704–11710. https://doi.org/10.1021/acsami.6b16559.

(27) Ma, Y.; Goodwill, J.; Skowronski, M. Quantification of Compositional Runaway during Electroformation in TaOx Resistive Switching Devices. *2019 IEEE 11th Int. Mem. Work. IMW 2019* **2019**, 6–9. https://doi.org/10.1109/IMW.2019.8739727.

(28) Funck, C.; Menzel, S.; Aslam, N.; Zhang, H.; Hardtdegen, A.; Waser, R.; Hoffmann-Eifert, S. Multidimensional Simulation of Threshold Switching in NbO2 Based on an Electric Field Triggered Thermal Runaway Model. *Adv. Electron. Mater.* **2016**, *2*, 1600169. https://doi.org/10.1002/aelm.201600169.

(29) Chen, H.; Millis, A.; Yi, D.; Lu, N. Rare-Earth Nickelates R NiO 3 : Thin Films and Heterostructures. **2018,** 81 046501. https://doi.org/10.1088/1361-6633/aaa37a.

(30) Del Valle, J.; Rocco, R.; Domínguez, C.; Fowlie, J.; Gariglio, S.; Rozenberg, M. J.; Triscone, J. M. Dynamics of the Electrically Induced Insulator-To-Metal Transition in Rare-Earth Nickelates. *Phys. Rev. B* **2021**, *104* (16), 1–7. https://doi.org/10.1103/PhysRevB.104.165141.

(31) Zhang, H. T.; Park, T. J.; Islam, A. N. M. N.; Tran, D. S. J.; Manna, S.; Wang, Q.; Mondal, S.; Yu, H.; Banik, S.; Cheng, S.; Zhou, H.; Gamage, S.; Mahapatra, S.; Zhu, Y.; Abate, Y.; Jiang, N.; Sankaranarayanan, S. K. R. S.; Sengupta, A.; Teuscher, C.; Ramanathan, S.

Reconfigurable Perovskite Nickelate Electronics for Artificial Intelligence. *Science (80-. ).* **2022**, *375* (6580), 533–539. https://doi.org/10.1126/science.abj7943.

(32) Guo, Q.; Farokhipoor, S.; Magén, C.; Rivadulla, F.; Noheda, B. Tunable Resistivity Exponents in the Metallic Phase of Epitaxial Nickelates. *Nat. Commun.* **2020**, *11*, 2949. https://doi.org/10.1038/s41467-020-16740-5.

(33) Guo, Q.; Noheda, B. From Hidden Metal-Insulator Transition to Planckian-like Dissipation by Tuning the Oxygen Content in a Nickelate. *npj Quantum Mater.* **2021**, *6*, 72. https://doi.org/10.1038/s41535-021-00374-x.

(34) Kato, G.; Hayashi, T.; Takeuchi, M. The all-or-none principle at the stimulated point of normal and narcotised nerve. *American Journal of Physiology*. **1928**, *83* (2), 690–697. https://doi.org/10.1152/ajplegacy.1928.83.2.690.

(35) Seyed-Ali Sadegh-Zadeh, C. K. All-or-None Principle and Weakness of HodgkinHuxley Mathematical Model. *Int. J. Biotechnol. Bioeng.* **2017**, *11*, 486-490. https://doi.org/10.5281/zenodo.1314592.

(36) J Caprio, C D Derby. Aquatic Animal Models in the Study of Chemoreception. *The Senses: A Comprehensive Reference.* **2008**, 97–133. https://doi.org *10.1016/b978-012370880-9.00073-6*.

(37) Lee, Y. W.; Kim, B. J.; Lim, J. W.; Yun, S. J.; Choi, S.; Chae, B. G.; Kim, G.; Kim, H. T. Metal-Insulator Transition-Induced Electrical Oscillation in Vanadium Dioxide Thin Film. *Appl. Phys. Lett.* **2008**, *92* (16). https://doi.org/10.1063/1.2911745.

(38) Das, S. K.; Nandi, S. K.; Marquez, C. V.; Rúa, A.; Uenuma, M.; Puyoo, E.; Nath, S. K.; Albertini, D.; Baboux, N.; Lu, T.; Liu, Y.; Haeger, T.; Heiderhoff, R.; Riedl, T.; Ratcliff, T.; Elliman, R. G. Physical Origin of Negative Differential Resistance in V3O5 and Its Application as a Solid-State Oscillator. *Adv. Mater.* **2022**,35,202208477. https://doi.org/10.1002/adma.202208477.

(39) Brown, T. D.; Bohaichuk, S. M.; Islam, M.; Kumar, S.; Pop, E.; Williams, R. S. Electro-Thermal Characterization of Dynamical VO2 Memristors via Local Activity Modeling. *Advanced Materials*. **2022**, 202205451. https://doi.org/10.1002/adma.202205451.

(40) Salverda, M.; Hamming-Green, R. P.; Noheda, B. An Epitaxial Perovskite as a Compact Neuristor: Electrical Self-Oscillations in TbMnO3 thin Films. *J. Phys. D. Appl. Phys.* **2022**, *55* (33). https://doi.org/10.1088/1361-6463/ac71e2.

(41) Kishida, H.; Ito, T.; Nakamura, A.; Takaishi, S.; Yamashita, M. Current Oscillation

Originating from Negative Differential Resistance in One-Dimensional Halogen-Bridged Nickel Compounds. *J. Appl. Phys.* **2009**, *106* (1), 1–4. https://doi.org/10.1063/1.3157211.

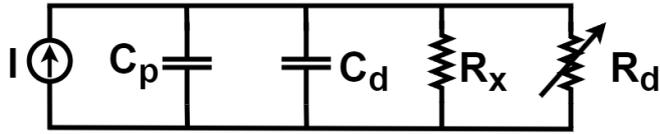

Fig. S2: Electrical circuit for current-control I-V plots.

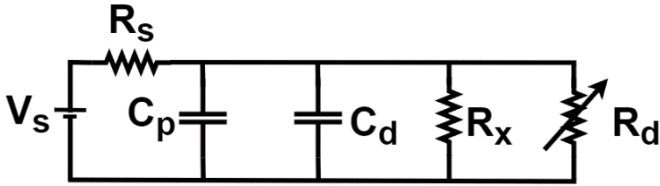

Fig. S3: Electrical circuit for Current oscillations.

The main strategy behind the model is to use only measured data for electrical conductivity vs temperature [add citation] to model the temperature dependent resistance and the rest of the physics is captured in lumped model for electrical and thermal components. The DC I-V characteristics were first reproduced and subsequently different sourcing condition that matched the measurements was applied to the model to reproduce the oscillatory behavior.
In the simulated circuit shown in Fig. S1, current is applied to the device, and the voltage across the device is measured.
In the simulated circuit illustrated in Fig. S2, load resistor Rs is connected between the voltage source and device to operate the device in NDR region.
In Fig. S1 and S2, $C_p$ and $C_d$ are the parasitic and device capacitance, respectively. $R_d$ is the device resistance, representative of the filament, that changes as the temperature across the device changes. $R_x$ is the fixed resistance that is not involved in filament formation and is not subject to significant temperature change as verified by our thermal model.
The values of Cp, Cd and width of filament formation are selected such that the simulated output for a single applied voltage is consistent with the experimental results. These values are held constant for all other input conditions.
The thermal circuit employed by us is shown in Figs. S3a and S3b. The output power Q (I($R_d$) * $V_d$) is fed to the thermal circuit marked as node A in Figs. S3a and S3b. Node A is in the middle of the channel of NNO. Heat can flow along three principal directions viz. i) along the channel length (Resistances with $RN_1$) as shown in Fig. 3a ii) along the thickness towards the bottom substrate (Resistances with $RN_{V1}$) as shown in Fig. 3a and iii) along the width outside the channel (Resistances with $RN_{3D}$) as shown in Fig. S3b.

Notation of thermal resistance and thermal capacitance with subscript 1 is the active area of the device(channel) and with subscript 2 is the area outside the channel. To maintain the symmetry in the circuit, capacitors are added at the middle of two thermal resistors as per standard practice.

Since the ambient temperature of the sample was simulated using a DC voltage source whose output was a voltage equal to ambient temperature. In particular, the illustrated model has a 10V DC voltage source representing ambient temperature of 10K. The output voltage at A in Fig.S3a corresponds to the temperature rise in the device. (Fig. 4d).

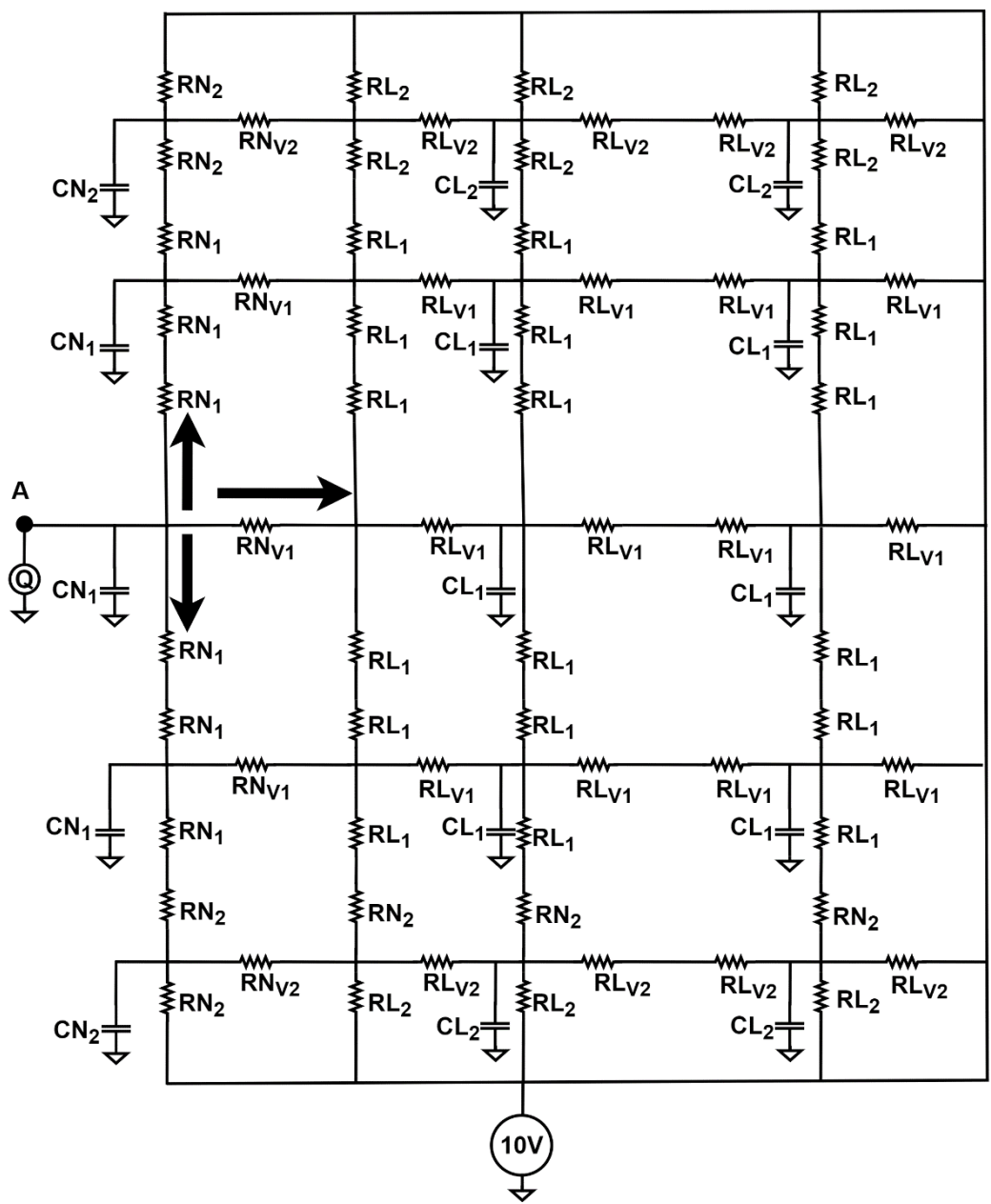

Figure S4a: Thermal Circuit. In resistors, N represents NNO. L represents LAO.

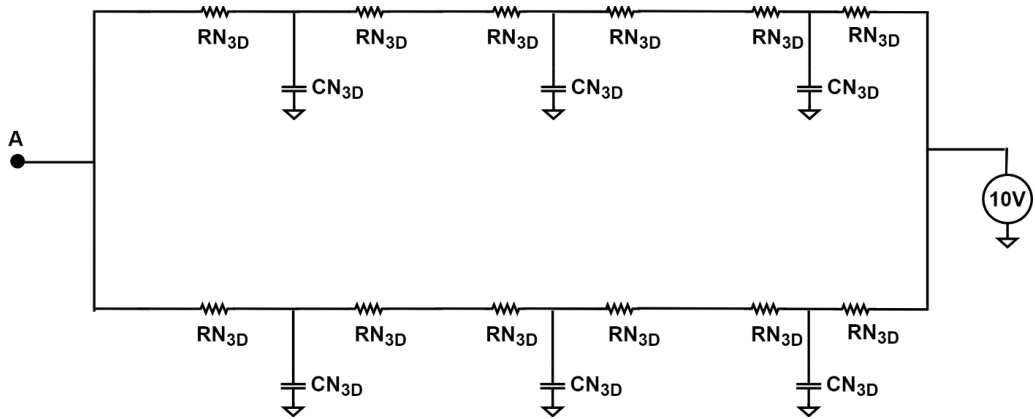

Figure S5b: Thermal circuit with resistors along the width of the channel.

| Parameters | Value | Reference |
|---|---|---|
| $\sigma_{ins}$ | 10 S/m | 1 |
| $\sigma_m$ | 2e6 S/m | 1 |
| $T_s$ | 120 K | 1 |
| $\alpha$ | 28 | Estimated |
| $\kappa_{ins}$ | 1 W/m/K | 2 |
| $\kappa_m$ | 3 W/m/K | 2 |
| $\rho$ | 7500 kg m$^3$ | |
| C_NNO | 2 J/kg/K | 3 |
| C_LAO | 141 J/kg/K | 4 |
| $C_d$ | 0.5e-9 F | Estimated |
| Channel length | 760e-9 m | SEM |
| Channel width | 5e-6 m | SEM |
| Filament width | 0.09e-6 m | |
| Sample Length | 5e-3 m | |
| Sample Width | 5e-3 m | |
| Thickness NNO | 5e-9 m | |
| Thickness LAO | 500e-6 m | |

**References:**

(1) Guo, Q.; Farokhipoor, S.; Magén, C.; Rivadulla, F.; Noheda, B. Tunable Resistivity Exponents in the Metallic Phase of Epitaxial Nickelates. *Nat. Commun.* **2020**, *11* (1), 1–9. https://doi.org/10.1038/s41467-020-16740-5.

(2) Hooda, M. K.; Yadav, C. S. Electronic Properties and the Nature of Metal-Insulator

Transition in NdNiO3 Prepared at Ambient Oxygen Pressure. *Phys. B Condens. Matter* **2016**, *491*, 31–36. https://doi.org/10.1016/j.physb.2016.03.014.

(3) Barbeta V.B., Jardim R.F., Torikachvili M.S., Escote M.S., Cordero F., Pontes F.M. Trequattrini F.Metal-Insulator Transition in Nd1−xEuxNiO3 Probed by Specific Heat and Anelastic Measurement *J. Appl. Phys.* **2011**, 109(7). https://doi.org/10.1063/1.3549615

(4) Michael, P. C.; Trefny, J. U.; Yarar, B. Thermal Transport Properties of Single Crystal Lanthanum Aluminate. *J. Appl. Phys.* **1992**, *72* (1), 107–109. https://doi.org/10.1063/1.352166.